\documentstyle[aps,preprint,floats,tighten,epsf]{revtex}

\begin{document}
\preprint{\vbox{\hbox{TRI--PP--96--60}
                \hbox{hep-lat/9610027}}}

\title{Light vector meson decay constants and the renormalization factor from a
       tadpole-improved action}

\author{Randy Lewis and R. M. Woloshyn}
\address{TRIUMF, 4004 Wesbrook Mall, Vancouver, British Columbia, Canada 
         V6T 2A3}

\date{October 1996}
\maketitle

\begin{abstract}
The $\rho$, $K^*$ and $\phi$ decay constants and the vector current
renormalization factor are studied by using
an ${\cal O}(a^2)$ classically-improved, tadpole-improved action.
Tree-level calculations are used to show how the classical
improvement of the action, involving next-nearest-neighbour timesteps, is 
transferred to the matrix elements.
Simulations are performed on coarse lattices and compared to Wilson
results from both coarse and fine lattices.  The improved action data are
found to resemble Wilson data obtained at 1/3 of the lattice spacing, which
is the same degree of improvement that is seen by comparing the mass spectra.
\end{abstract}


\section{INTRODUCTION}

Numerical simulations of QCD can be performed by discretizing Euclidean
space-time onto a lattice and working within a finite volume.
Physical measurements are then obtained by extrapolating to
zero lattice spacing and infinite lattice volume.  There are many lattice 
actions which correspond to continuum QCD in the appropriate limit, and a
lot of recent work has focussed on finding actions which reduce 
the errors due to discrete lattice spacing. The goal is to find actions
which provide accurate results on coarsely-spaced lattices
and so allow lattice sizes (and therefore computing resources) to be
reduced. A progress report on improving actions may be found 
in Ref.~\cite{Nie}.

To demonstrate the value of an improved action, one would like to show that
accurate results can be obtained from coarse lattices
for a wide variety of quantities.  As a first step beyond the mass spectrum, 
where significant improvements on coarse lattices have already been
demonstrated, we consider 
the decay constants of light vector mesons and the associated
vector current renormalization factor.  
(A preliminary discussion was given in 
Ref.~\cite{LewWol}.)  An advantage of studying vector currents is the 
existence of
a conserved vector current even at nonzero lattice spacing, which does not
get renormalized.  Neglecting lattice spacing errors, the matrix element of 
the conserved current between any initial and final states is identical to 
the matrix element of the local vector current between those same states, 
except for the multiplicative ``renormalization factor'', $Z_V$.  In this way,
$Z_V$ can be easily computed.
In practice, it has been known for a long time\cite{HeaSac} that the 
determination of $Z_V$ is quite sensitive to lattice spacing errors, which
is another reason that the vector current is valuable for the present work,
since it is precisely these errors that should be reduced by an improved
action.

The action\cite{LusWei85,Hametc} that will be used here is 
improved at the classical level up to and including ${\cal O}(a^2)$, 
and is also improved at the quantum level 
via the ``tadpole factor'' of Lepage and Mackenzie\cite{LepMac}.
The action has the added feature of being quite simple, since the gauge sector
contains only $1 \times 1$ and $1 \times 2$ plaquettes, and the fermion sector
does not contain the so-called ``clover'' term and has no need of the 
associated field transformations.

The mass spectrum of this improved
action has been studied numerically in Ref.~\cite{FieWol}, 
where it was concluded
that data from the original Wilson action\cite{Wil} at a lattice spacing of
$a \sim 0.1$fm can be reproduced with the improved
action on lattices with $a \sim 0.3$fm, representing a very substantial saving 
of computing resources.
Other studies of the mass spectrum with the same improved action have produced 
similar results\cite{LeeLei,BordeF}. 

In this paper it is shown that
a comparable degree of improvement is also obtained in the computation of 
the vector meson decay constants, $f_\rho$ , $f_{K^*}$ and $f_\phi$, and the 
renormalization factor, $Z_V$.  Sec.~\ref{Sec:method} begins with a review of
the general methods for calculating vector current matrix elements, 
and then discusses
the particular improved action chosen for this work.
Sec.~\ref{Sec:treelevel} is used to show explicitly that the
quantities to be computed are classically improved.  The results of numerical
simulations using the improved action are compared to Wilson action results
in Sec.~\ref{Sec:simulation}, and a summary is given
in Sec.~\ref{Sec:conclude}.

\section{METHOD OF CALCULATION}\label{Sec:method}

The $\rho$ meson decay constant, $f_\rho$, is defined from a continuum matrix
element of the local vector current as follows.
\begin{equation}\label{definefV}
   \langle 0|V_\mu^L|\rho\rangle_{\rm cont} =
   \frac{m_\rho^2}{f_\rho}\epsilon_\mu
\end{equation}
\begin{equation}
   V_\mu^L(x) = \bar{\psi}(x)\gamma_\mu\psi(x)
\end{equation}
A similar definition applies to $f_{K^*}$ and $f_\phi$.  Recall that this 
current is conserved in the continuum.

At nonzero lattice spacing the local current is not conserved, and
its matrix element between any initial and final states
gets renormalized by a factor $Z_V$,
\begin{equation}
   \langle f|V_\mu^L|i\rangle_{\rm cont} =
               Z_V(g^2)\langle f|V_\mu^L|i\rangle + {\cal O}(a^n)~.
\end{equation}
$g$ is the lattice coupling, and ${\cal O}(a^n)$ represents the contributions
from operators of 
higher mass dimension which are multiplied by powers of the lattice spacing, 
$a$.  For the Wilson action these corrections begin at ${\cal O}(a)$, but for 
our improved action no
${\cal O}(a,a^2)$ classical terms can appear, and the tadpole factor is 
expected to approximately minimize terms proportional to the coupling, for 
example ${\cal O}(g^2a)$ terms.

If all quark flavours have the same mass, then both the improved action and
the Wilson action have an exact vector (flavour) symmetry, which implies the 
existence of a conserved ``Noether'' current, $V_\mu^C$.
(Actually, there is not a unique conserved current, and care must be
used to choose one that is classically improved to the appropriate order.
This will be addressed below.)  
Because of the nonrenormalization of conserved currents,
\begin{equation}
   \langle f|V_\mu^L|i\rangle_{\rm cont} =
               \langle f|V_\mu^C|i\rangle + {\cal O}(a^n)~,
\end{equation}
which allows $Z_V$ to be computed from a ratio of matrix elements as follows.
\begin{equation}\label{ratio}
   Z_V(g^2) =
   \frac{\langle f|V_\mu^C|i\rangle}{\langle f|V_\mu^L|i\rangle} 
   + {\cal O}(a^n)
\end{equation}
Although $Z_V$ itself is independent of the initial and final states, the 
ratio of matrix elements in Eq.~(\ref{ratio}) is not, so the size of the 
lattice spacing errors can be estimated by varying $|i\rangle$ and $|f\rangle$.
In this work, we will consider four separate choices for the ratio in
Eq.~(\ref{ratio}): a vector current insertion between pseudoscalar meson, 
vector meson or spin-1/2 baryon states,
\begin{equation}
   R_P \equiv \frac{\langle P|V_\mu^C|P\rangle}{\langle P|V_\mu^L|P\rangle}
   ~~~,~~~
   R_V \equiv \frac{\langle V|V_\mu^C|V\rangle}{\langle V|V_\mu^L|V\rangle}
   ~~~,~~~
   R_B \equiv \frac{\langle B|V_\mu^C|B\rangle}{\langle B|V_\mu^L|B\rangle} 
\end{equation}
and the decay process,
\begin{equation}
   R_D \equiv \frac{\langle 0|V_\mu^C|V\rangle}{\langle 0|V_\mu^L|V\rangle}~.
\end{equation}

In our lattice simulations, states will be created via the following local 
interpolating fields.
\begin{eqnarray}\label{interpol1}
   {\rm pseudoscalar~mesons:~}\chi^P(x) &=& \bar{\psi}(x)\gamma_5\psi(x) \\
   {\rm vector~mesons:~}\chi^V_\mu(x) &=& \bar{\psi}(x)\gamma_\mu\psi(x) \\
   {\rm spin-}\frac{1}{2}{\rm ~baryons:~}\chi^B(x) &=& 
                     \epsilon_{abc}\psi_a(x)\left[\psi^T_b(x)C\gamma_5
                     \psi_c(x)\right] \label{interpol2}
\end{eqnarray}
($C$ is the charge conjugation matrix.)  
The decay constant itself is obtained from simulations of the two-point vector
correlator,
\begin{equation}\label{calcfV}
   \sum_{\bf x}\sum_{\mu=1,3}\langle V^L_\mu({\bf x},t)V_\mu^{L\,\dagger}(0)
   \rangle \rightarrow \frac{3m_V^3e^{-m_Vt}}{2(2\kappa)^2Z_V^2f_V^2} 
   ~~~,~~{\rm for~}t \gg 0,
\end{equation}
which is derived by multiplying Eq.~(\ref{definefV}) with its hermitian 
conjugate, converting from continuum to lattice quantities, and performing the 
appropriate summations.  The explicit dependence on $Z_V$ can be changed by
replacing one (or both) of the local currents in Eq.~(\ref{calcfV})
by a conserved current.
On a lattice with periodic boundary conditions, another term must be added to
the right-hand side of Eq.~(\ref{calcfV}) which decays exponentially from
the (periodic) image of the source.

Following the work of L\"uscher and Weisz\cite{LusWei85}, 
the gauge field term of the improved action involves a sum over $1 \times 2$
rectangular plaquettes ($U_{rt}$) as well as $1 \times 1$ elementary plaquettes 
($U_{pl}$),
\begin{equation}
   S_G(U) = \frac{\beta}{3}{\rm ReTr}\left[\sum_{pl}(1-U_{pl})
        -\frac{1}{20U_0^2}\sum_{rt}(1-U_{rt})\right]~,
\end{equation}
where a tadpole factor, defined by
\begin{equation}
   U_0 = \langle\frac13\mbox{Re}\mbox{Tr}U_{pl}\rangle^{1/4}~,
\end{equation}
has been introduced to absorb the lattice tadpole effects and thereby improve
the matching to perturbation theory\cite{LepMac}.
The fermion part,
which was first introduced more than a decade ago\cite{Hametc}, 
also contains next-nearest-neighbour interactions.
\begin{eqnarray}\label{S_F}
   S_F(\bar{\psi},\psi;U) &=& -\sum_{x}\bar{\psi}(x)\psi(x) \nonumber \\
        && +\frac{4}{3}\kappa\sum_{x,\mu}\left[
           \bar{\psi}(x)(1-\gamma_{\mu})U_{\mu}(x)\psi(x+\mu)
           +\bar{\psi}(x+\mu)(1+\gamma_{\mu})
           U_{\mu}^{\dagger}(x)\psi(x)\right] \nonumber \\
        && -\frac{\kappa}{6U_0}\sum_{x,\mu}\left[\bar{\psi}(x)(2-\gamma_{\mu})
           U_{\mu}(x)U_{\mu}(x+\mu)\psi(x+2\mu) \right. \nonumber \\
        && \phantom{-\frac{\kappa}{6U_0}\sum_{x,\mu}}\left.
           +\bar{\psi}(x+2\mu)(2+\gamma_{\mu})
           U_{\mu}^{\dagger}(x+\mu)U_{\mu}^{\dagger}(x)\psi(x)\right]
\end{eqnarray}
The presence of next-nearest-neighbour timesteps is a generic consequence of 
classical improvement beyond ${\cal O}(a)$, and generates well-known 
artifacts, e.g. unphysical branches in the free dispersion 
relations, discussed for gauge fields in Ref.~\cite{LusWei84}.
The effect is also seen for fermions, as discussed in following sections.

The Noether vector current corresponding to Eq.~(\ref{S_F}) is
\begin{eqnarray}
   V_\mu^C(x) &=& \frac{2}{3}
        \bar{\psi}(x+\mu)(1+\gamma_{\mu})U_{\mu}^{\dagger}(x)\psi(x)
    -\frac{2}{3}\bar{\psi}(x)(1-\gamma_{\mu})U_{\mu}(x)\psi(x+\mu) \nonumber \\
    && +\frac{1}{12U_0}\bar{\psi}(x)(2-\gamma_{\mu})
           U_{\mu}(x)U_{\mu}(x+\mu)\psi(x+2\mu) \nonumber \\
    && +\frac{1}{12U_0}\bar{\psi}(x-\mu)(2-\gamma_{\mu})
           U_{\mu}(x-\mu)U_{\mu}(x)\psi(x+\mu) \nonumber \\
    && -\frac{1}{12U_0}\bar{\psi}(x+2\mu)(2+\gamma_{\mu})
           U_{\mu}^{\dagger}(x+\mu)U_{\mu}^{\dagger}(x)\psi(x) \nonumber \\
    && -\frac{1}{12U_0}\bar{\psi}(x+\mu)(2+\gamma_{\mu})
           U_{\mu}^{\dagger}(x)U_{\mu}^{\dagger}(x-\mu)\psi(x-\mu)~.
\end{eqnarray}
Although this current is conserved, an explicit calculation shows that
it is not classically improved to the level of the action because of extra
total derivative terms.
An improved conserved current could be written as follows\cite{Lep},
\begin{equation}
   V_\mu^{CI}(x) = \frac{12}{13}V_\mu^C(x) + \frac{1}{26}V_\mu^C(x-\mu)
                 + \frac{1}{26}V_\mu^C(x+\mu)~,
\end{equation}
but the offending total derivatives in $V_\mu^C$ do not contribute to the
quantities we wish to compute, as will be shown explicitly in 
Sec.~\ref{Sec:treelevel}, so $V_\mu^{CI}$ is not required for this work.

\section{CONFIRMING TREE-LEVEL IMPROVEMENT}\label{Sec:treelevel}

The matrix elements that are used to obtain $f_\rho$, $f_{K^*}$, $f_\phi$ and 
$Z_V$ can be
evaluated analytically at the classical level.  We will follow closely the
approach and notation of El-Khadra, Kronfeld and 
Mackenzie\cite{ElKKroMac}, paying particular attention to the effects of 
next-nearest-neighbour interactions, which were not present in the actions of 
Ref.~\cite{ElKKroMac}.

\subsection{The quark equation of motion}

\begin{figure}[p]
\epsfxsize=380pt \epsfbox[30 419 498 732]{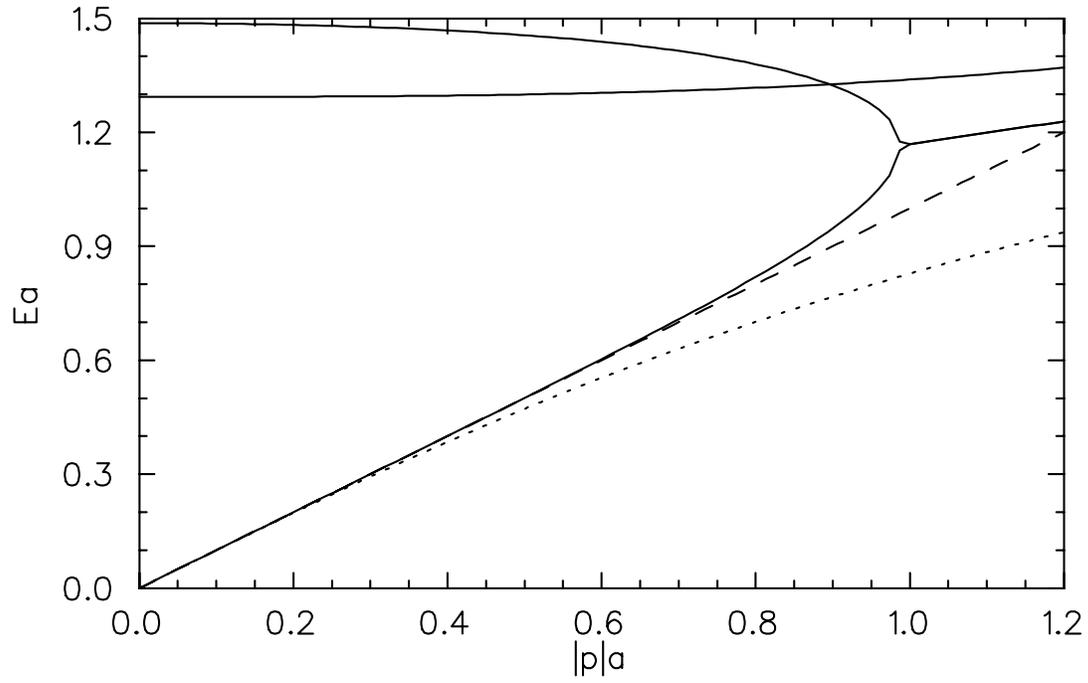}
\vspace{18pt}
\caption{The energy-momentum relation for a massless, free fermion from the 
         improved 
         action with 3-momentum $\bf p$ in the (1,1,0) direction.  Only the 
         real parts of complex solutions are shown.  The dashed(dotted) curve 
         is the continuum(Wilson) result.
         }\label{fig:Evsp}
\end{figure}

The equation of motion that comes from Eq.~(\ref{S_F}) can be written in the
form
\begin{equation}\label{EOM}
   \sum_{y}\left[\sum_\mu\gamma_{\mu}\tilde{K}_\mu(x,y)+\tilde{L}(x,y)
   \right]\psi(y) = 0
\end{equation}
where the Fourier transforms of $\tilde{K}_\mu(x,y)$ and $\tilde{L}(x,y)$ are
\begin{eqnarray}
  iK_\mu(p) &=& \frac{i}{3}{\rm sin}p_\mu{a}\left(4-{\rm cos}p_\mu{a}\right) \\
   L(p) &=& Ma+4-\frac{4}{3}\sum_\mu{\rm cos}p_\mu{a}
            +\frac{1}{3}\sum_\mu{\rm cos}2p_\mu{a}
\end{eqnarray}
respectively.  The on-shell condition is
\begin{equation}\label{KKLL}
   \sum_\mu{K}_\mu^2(p) + L^2(p) = 0
\end{equation}
which is a quartic equation in ${\rm cosh}Ea = {\rm cos}p_4a$.
Fig.~\ref{fig:Evsp} shows the solutions for 3-momenta in the
(1,1,0) direction and $M=0$, compared to the Wilson action result and the 
continuum $E=|{\bf p}|$ line.  Notice that the improved action has artifacts 
which are restricted to the region of large $Ea$, and that the low-energy 
solution is closer to the
continuum for a greater range of $|{\bf p}|a$ than is the Wilson curve.

\begin{figure}[p]
\epsfxsize=380pt \epsfbox[30 419 498 732]{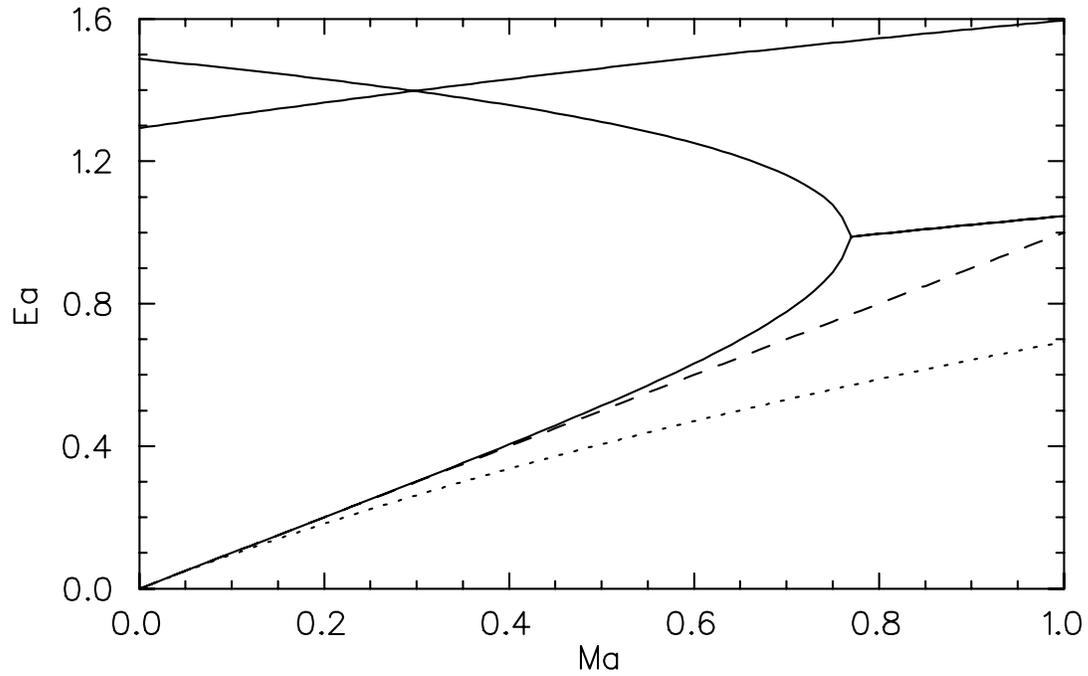}
\vspace{18pt}
\caption{The energy-mass relation for a stationary, free fermion from the 
         improved action.  Only the real
         parts of complex solutions are shown.  The dashed(dotted) curve is
         the continuum(Wilson) result.
         }\label{fig:EvsM}
\end{figure}

Also related to our discussion of decay constants is the limit of
vanishing 3-momentum, ${\bf p}=0$, and Fig.~\ref{fig:EvsM} shows the solutions
of Eq.~(\ref{KKLL}) versus $Ma$.  Again, improvement is found with respect to
the Wilson action.  In this case as well there are next-nearest-neighbour 
timestep artifacts which, however, remain at
high energies for the values of $Ma$ that we will use in numerical simulations.
Note that Eq.~(\ref{KKLL}) is only quadratic in the mass, and becomes the
following unique expression for $Ma$ when ${\bf p} = {\bf 0}$.
\begin{equation}\label{useful}
   Ma+1 = \frac{1}{3}\left[1+4{\rm cosh}Ea-2{\rm cosh}^2Ea+(4-{\rm cosh}Ea)
          {\rm sinh}|Ea|\right]
\end{equation}

We now seek spinor solutions to the equations of motion,
\begin{eqnarray}
  \left[i\sum_\mu\gamma_\mu{K}_\mu(p)+L(p)\right]u(\xi,{\bf p}) &=& 0 \\
  \left[-i\sum_\mu\gamma_\mu{K}_\mu(p)+L(p)\right]v(\xi,{\bf p}) &=& 0
\end{eqnarray}
where $\xi = 1,2$ is the spin index.
The equation of motion for $v$-spinors has been modified so that from now
on $E$ is always a positive quantity.
The solutions are
\begin{eqnarray}
   u(\xi,{\bf p}) &=& \left[\frac{-i\gamma_\mu{K}_\mu(p)
      +L(p)}{\sqrt{2L(p)\left[L(p)+{\rm sinh}Ea\left(4-{\rm cosh}Ea
      \right)/3\right]}}\right]u(\xi,{\bf 0}) \\
   v(\xi,{\bf p}) &=& \left[\frac{i\gamma_\mu{K}_\mu(p)
      +L(p)}{\sqrt{2L(p)\left[L(p)+{\rm sinh}Ea\left(4-{\rm cosh}Ea
      \right)/3\right]}}\right]v(\xi,{\bf 0})
\end{eqnarray}
where
\begin{equation}
   u(1,{\bf 0}) = \left(\begin{tabular}{c}1\\0\\0\\0\end{tabular}\right)~~~,~~
   u(2,{\bf 0}) = \left(\begin{tabular}{c}0\\1\\0\\0\end{tabular}\right)~~~,~~
   v(1,{\bf 0}) = \left(\begin{tabular}{c}0\\0\\1\\0\end{tabular}\right)~~~,~~
   v(2,{\bf 0}) = \left(\begin{tabular}{c}0\\0\\0\\1\end{tabular}\right)~.
\end{equation}

The general solution of Eq.~(\ref{EOM}) is
\begin{equation}
   \psi(x) = \int\frac{{\rm d}^3{\bf p}}{(2\pi)^3}N({\bf p})\sum_\xi
             \left[b(\xi,{\bf p})u(\xi,{\bf p})e^{ip{\cdot}x}
             +d^\dagger(\xi,{\bf p})v(\xi,{\bf p})e^{-ip{\cdot}x}\right]
\end{equation}
where $b(\xi,{\bf p})$ annihilates a quark and $d^\dagger(\xi,{\bf p})$ 
creates an antiquark.
The normalization factor $N({\bf p})$ can be determined, for example, from the
normalization of the quark propagator.  By choosing the creation and 
annihilation operators to be normalized in the following way,
\begin{equation}
   \left\{b(\xi^\prime,{\bf p}^\prime),b^\dagger(\xi,{\bf p})\right\} =
   \left\{d(\xi^\prime,{\bf p}^\prime),d^\dagger(\xi,{\bf p})\right\} =
   (2\pi)^3\delta^3({\bf p}^\prime-{\bf p})\delta^{\xi\xi^\prime}~,
\end{equation}
the normalization factor is found to be
\begin{equation}
   N({\bf p}) = \left[\frac{9L(p)}{\left(4{\rm cosh}Ea-{\rm cosh}2Ea\right)
                {\rm sinh}Ea\left(4-{\rm cosh}Ea\right)+6\left(2{\rm sinh}Ea
                -{\rm sinh}2Ea\right)L(p)}\right]^{1/2}~.
\end{equation}

\subsection{$f_V$ and $Z_V$}

The vector meson decay amplitude 
and the insertion of a vector current between hadron states
can now be calculated at the classical level using the matrix elements
\[
   \langle0|V_\mu|q(\xi,{\bf p})\bar{q}(\xi^\prime,{\bf p^\prime})\rangle
   ~~~{\rm and}~~~
   \langle{q}(\xi^\prime,{\bf p^\prime})|V_\mu|q(\xi,{\bf p})\rangle
\]
respectively.
In each case, kinematics are chosen that corresponds to the numerical
simulations of Sec.~\ref{Sec:simulation}, which means that
the decay amplitude is studied in the rest frame (${\bf p^\prime}=-{\bf p}$),
whereas the vector current insertion has no 4-momentum transfer 
($p^\prime = p$).
A consequence of these choices is that only $\mu=1,2,3$ are nonzero for the
decay amplitude, and only $\mu=4$ is nonzero for the vector current insertion.

Using $V^C_\mu$ as the vector current, the decay amplitude is
\begin{eqnarray}
   \langle0|V^C_j|q(\xi,{\bf p})\bar{q}(\xi^\prime,-{\bf p})\rangle
   \!\!\!\!\!\!\!\!\!\!\!\!\!\!\!\!\!\!\!\!\!\!\!\!\!\!\!\!\!\!\!\!\!
   \!\!\!\!\! && \nonumber \\
   &=& N^2({\bf p})\bar{v}(\xi^\prime,-{\bf p})\left[-\frac{4i}{3}{\rm sin}
       p_ja+\frac{4}{3}\gamma_j{\rm cos}p_ja+\frac{2i}{3}{\rm sin}2p_ja
       -\frac{1}{3}\gamma_j{\rm cos}2p_ja\right]u(\xi,{\bf p}) \\
   &=& N^2({\bf p})\left[\frac{4}{3}{\rm cos}p_ja-\frac{1}{3}{\rm cos}2p_ja
       \right]\bar{v}(\xi,{\bf 0})\gamma_ju(\xi,{\bf 0}) \\
   &=& \bar{v}(\xi,{\bf 0})\gamma_ju(\xi,{\bf 0})\left[1+{\cal O}(a^3)\right]
       \label{classdecay}
\end{eqnarray}
where Eq.~(\ref{useful}) has been used.
A similar derivation shows that Eq.~(\ref{classdecay}) also holds for the local
vector current, $V^L_\j$, so the decay amplitude is classically 
improved up to ${\cal O}(a^2)$ for both vector currents.

The expression for the current insertion is
\begin{eqnarray}
   \langle{q}(\xi^\prime,{\bf p})|V^C_4|q(\xi,{\bf p})\rangle
   \!\!\!\!\!\!\!\!\!\!\!\!\!\!\!\!\!\!\!\!\!\!\!\! && \nonumber \\
   &=& N^2({\bf p})\bar{u}(\xi^\prime,{\bf p})\left[\frac{4}{3}{\rm sin}Ea
       +\frac{4}{3}\gamma_4{\rm cos}Ea-\frac{2}{3}{\rm sin}2Ea-\frac{1}{3}
       \gamma_4{\rm cos}2Ea\right]u(\xi,{\bf p}) \\
   &=& \delta^{\xi\xi^\prime}~.
\end{eqnarray}
Notice that there are no finite-$a$ errors here as required by current 
conservation.  For the local current, it is found that
\begin{eqnarray}
   \langle{q}(\xi^\prime,{\bf p})|V^L_4|q(\xi,{\bf p})\rangle
   &=& \delta^{\xi\xi^\prime}\left[1+{\cal O}(a^3)\right]
\end{eqnarray}
which again displays the desired classical improvement.

Thus, although we are using an {\it un\/}improved conserved current, the
particular calculations being considered in this work are still improved to
the level of the action.  This conclusion will be valid for the numerical
simulations as well, provided we create the external hadron states using
operators that maintain the improvement.
Local interpolating fields, such as those of 
Eqs.~(\ref{interpol1}-\ref{interpol2}) which will be used in the present work,
guarantee that classical improvement is not destroyed.

\section{NUMERICAL SIMULATIONS}\label{Sec:simulation}

Calculations were performed at two values of $\beta$ with the improved
action, corresponding to lattice spacings of 0.4fm and 0.27fm as derived from
the string tension.  For comparison, the Wilson action was also used to
calculate at the same lattice spacings.  
Some details of the simulations are provided in Table~\ref{tab:details}.
\begin{table*}
\caption{Simulation parameters.  $N_U$ is the number of gauge field
         configurations, $a_{st}$ is the lattice spacing derived from the
         string tension, $\kappa_c$ is the hopping parameter at the critical 
         point and $\kappa_s$ is the hopping parameter corresponding
         to the strange quark mass (from $m_{K^*}/m_K$).  The error in the 
         last digit is bracketed.}\label{tab:details}
\begin{tabular}{ccccccc}
\rule[-2mm]{0mm}{6mm} Lattice & $N_U$ & $\beta$ & $a_{st}$[fm] & $\kappa$ &
                      $\kappa_c$ & $\kappa_s$ \\
\hline
\multicolumn{7}{c}{\rule[-2mm]{0mm}{8mm}Improved Action} \\
 $6^3$${\times}12$ 
         & 200 & 6.25 & 0.4  & 0.162, 0.168, 0.174 & 0.1803(2) & 0.1658(8) \\
 $8^3$${\times}14$ 
         & 100 & 6.8  & 0.27 & 0.150, 0.154, 0.158 & 0.1638(2) & 0.1561(5) \\
\multicolumn{7}{c}{\rule[-2mm]{0mm}{8mm}Wilson Action} \\
 $6^3$${\times}12$ 
         & 200 & 4.5  & 0.4  & 0.189, 0.201, 0.213 & 0.2189(2) & 0.2050(3) \\
 $8^3$${\times}14$ 
         & 100 & 5.5  & 0.27 & 0.164, 0.172, 0.180 & 0.1857(3) & 0.1779(7) \\
\end{tabular}
\end{table*}

In all cases, we used pseudo-heatbath
updating with periodic boundary conditions for the gauge fields.
The lattices were thermalized with 4000 sweeps, and then 250 sweeps (200
sweeps) were discarded between each pair of improved (Wilson) configurations
that was kept.  A stabilized biconjugate gradient algorithm was used for the
fermion matrix inversion.  To maximize the number of useful time slices,
Dirichlet boundary conditions were used for fermions in the time direction,
with periodic boundary conditions in all spacial directions.  
The source was placed
two timesteps away from the time boundary in every case, and for the
3-point correlators of $R_P$, $R_V$ and $R_B$ we attached the vector current
insertion to the lattice's centre time slice.

\begin{figure}[th]
\epsfxsize=380pt \epsfbox[30 419 498 732]{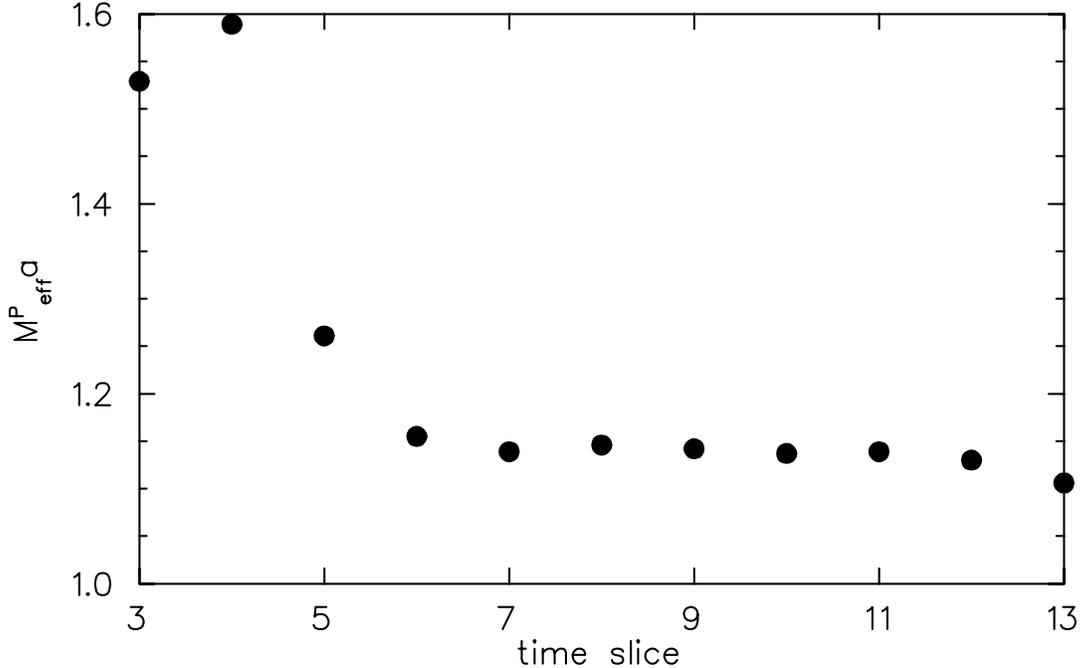}
\vspace{18pt}
\caption{The effective mass plot for a pseudoscalar meson using the improved 
         action at $\beta = 6.8$ with $\kappa=0.150$.
         }\label{fig:effmass}
\end{figure}

Fig.~\ref{fig:effmass} shows the effective mass plot for a pseudoscalar
meson with the improved action.  The non-monotonic behaviour near the source
($t=3$ in Fig.~\ref{fig:effmass})
is a consequence of next-nearest-neighbour interactions, as discussed
by L\"uscher and Weisz\cite{LusWei84}.
The effect of these high-energy oscillations decreases
as $\kappa$ increases, and Fig.~\ref{fig:effmass} shows that there is already
an identifiable plateau at our smallest value of $\kappa$.

To extract the mass from an effective mass plot such as Fig.~\ref{fig:effmass},
or to obtain $Z_V$ or $1/f_V$ from the appropriate plot, we require that the
plateau contains at least three neighbouring points, and we never use the
two time slices near the boundary (12 and 13 in Fig.~\ref{fig:effmass})
which can be affected by the Dirichlet boundary condition.  The systematic
uncertainties associated with choosing a plateau region will not be presented
since we do not expect these uncertainties to affect the principal conclusions
of our study.  All statistical errors will be computed using half the
distance between the 16th and 84th percentiles from 500 bootstrap samples.

\begin{figure}[p]
\epsfxsize=380pt \epsfbox[30 419 498 732]{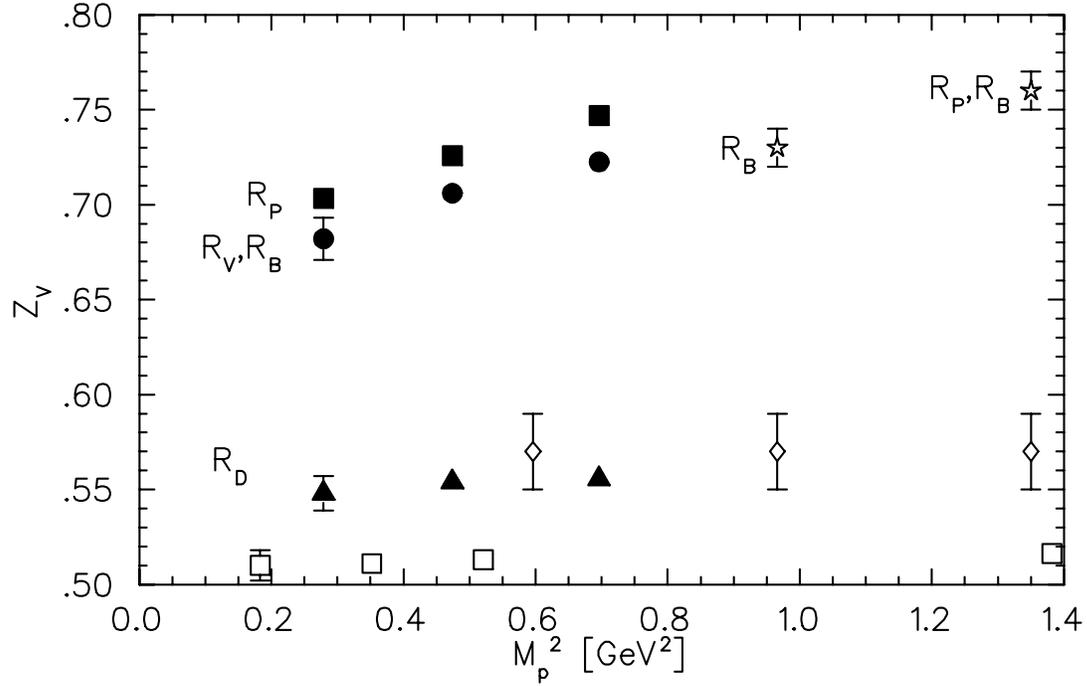}
\vspace{18pt}
\caption{The renormalization constant as a function of the pseudoscalar mass
         squared with degenerate quarks.  Solid symbols are from the improved
         action at $\beta = 6.8$; open stars\protect\cite{MarSac} and 
         diamonds\protect\cite{MaiMar} are Wilson at $\beta
         = 6.0$; open squares\protect\cite{QCDPAX} are Wilson at $\beta = 5.85$.
         }\label{fig:Zrawequ}
\end{figure}
The results for the renormalization factor, obtained from the improved action 
at $\beta=6.8$, are shown for degenerate-mass quarks in Fig.~\ref{fig:Zrawequ}.
The value of $Z_V$ that comes from using meson decay amplitudes ($R_D$) in 
Eq.~(\ref{ratio}) is significantly smaller than the values that come from
inserting a vector current between pion, $\rho$ meson or nucleon states
($R_P$, $R_V$, $R_B$),
and this difference is a reflection of the finite-$a$ errors that are present 
in Eq.~(\ref{ratio}).  To decide whether these finite-$a$ errors should be
considered as large or small, a comparison is made to results
from the Wilson action.
Fig.~\ref{fig:Zrawequ} shows that the improved action results
agree remarkably well with Wilson data at $\beta=6.0$, while being clearly
distinguishable from Wilson data (for $R_D$) at $\beta=5.85$.

\begin{figure}[p]
\epsfxsize=380pt \epsfbox[30 419 498 732]{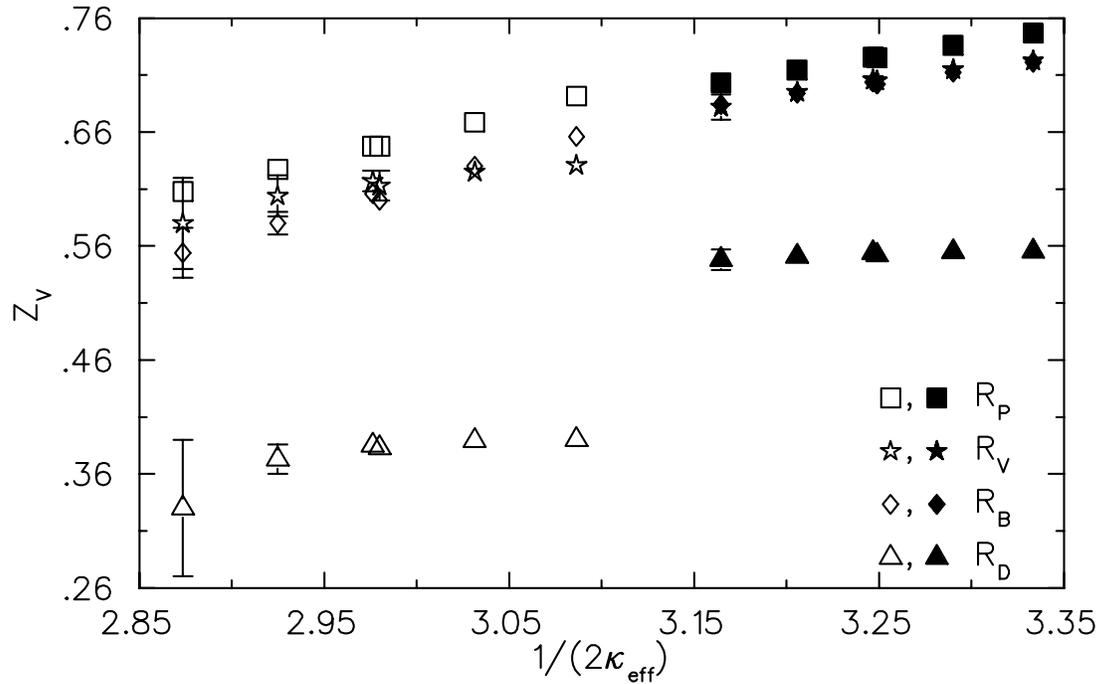}
\vspace{18pt}
\caption{The renormalization constant as a function of $1/(2\kappa_{\rm eff})$.
         Open(solid) symbols are from the improved action at 
         $\beta = 6.25(6.8)$.
         }\label{fig:Zrawtot}
\end{figure}
By defining an effective hopping parameter which averages over the two quark
flavours in a meson or baryon,
\begin{equation}\label{kapeff}
   \frac{1}{2\kappa_{\rm eff}} = \frac{1}{2}\left(\frac{1}{2\kappa_1}+
                                 \frac{1}{2\kappa_2}\right)~,
\end{equation}
we can plot all of our data, involving degenerate and nondegenerate
quarks, together.  Fig.~\ref{fig:Zrawtot} shows that our improved action
results for $Z_V$ appear to be linear in $1/(2\kappa_{\rm eff})$. 

\begin{figure}[p]
\epsfxsize=380pt \epsfbox[30 419 498 732]{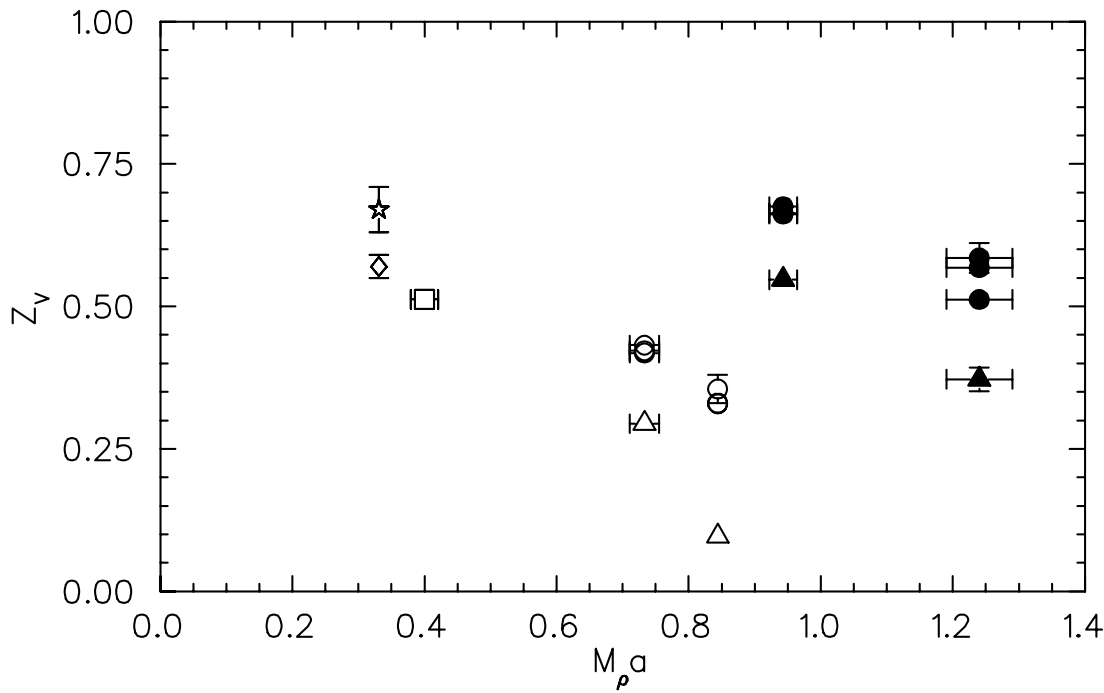}
\vspace{18pt}
\caption{The renormalization constant in the chiral limit.  Solid(open) 
         symbols are from the improved(Wilson) action.  The open star,
         diamond and square are from Refs.~\protect\cite{MarSac}, 
         \protect\cite{MaiMar} and \protect\cite{QCDPAX} respectively.  The 
         lowest-lying point at each $M_\rho{a}$ is from the decay amplitude,
         $R_D$.
         }\label{fig:Zlight}
\end{figure}

The chiral limit, $\kappa = \kappa_c$, is reached by extrapolating $Z_V$, 
$1/f_V$,
pseudoscalar meson masses {\it squared}, vector meson masses and baryon masses 
linearly in $1/\kappa$, using three data points (degenerate and nondegenerate
quarks are used separately).
In this way, we obtain the results of Fig.~\ref{fig:Zlight} for the 
renormalization factor without strange quark involvement.
Also shown are the chiral limit values corresponding
to the Wilson results of Fig.~\ref{fig:Zrawequ}, and our own Wilson
calculations at smaller $\beta$'s.  Note that the improved
action approximately resembles the trajectory of Wilson data, but translated 
horizontally to lattice spacings about three times as large. This is very
similar to what was also found in a comparison of the mass 
spectra\cite{FieWol}, and provides the first indication that the accuracy
of calculations with tadpole-improved actions on coarse lattices
persists beyond spectral quantities.

\begin{figure}[p]
\epsfxsize=380pt \epsfbox[30 419 498 732]{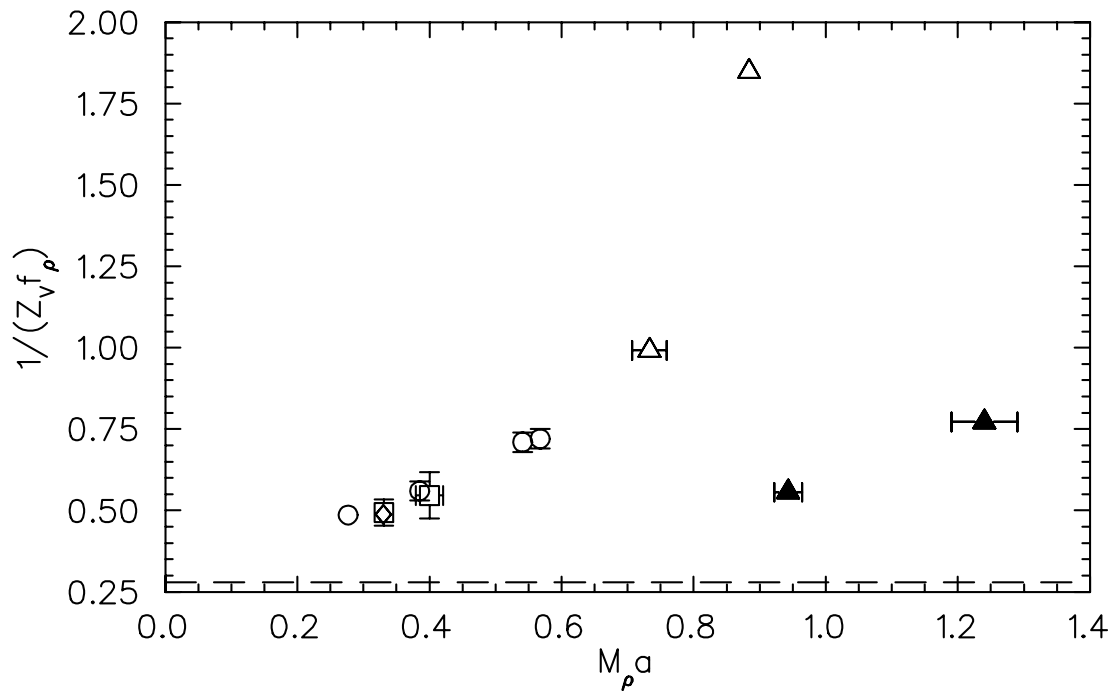}
\vspace{18pt}
\caption{The lattice decay constant in the chiral limit.  Solid(open) 
         symbols are from the improved(Wilson) action and the dashed line
         is the experimental value. Open squares, circles and diamonds are 
         from Refs.~\protect\cite{QCDPAX}, \protect\cite{GF11} and
         \protect\cite{BhaGup} respectively.
         All Wilson quarks are normalized by $\sqrt{2\kappa}$.
         }\label{fig:ZfV}
\end{figure}

Fig.~\ref{fig:ZfV} shows the $\rho$ meson decay constant without including the 
effect of $Z_V$, and with all Wilson quarks normalized by $\sqrt{2\kappa}$.
Again, it is found that the improved action gives results which resemble the
Wilson data, but at a lattice spacing about three times as large.

\begin{figure}[p]
\epsfxsize=380pt \epsfbox[30 419 498 732]{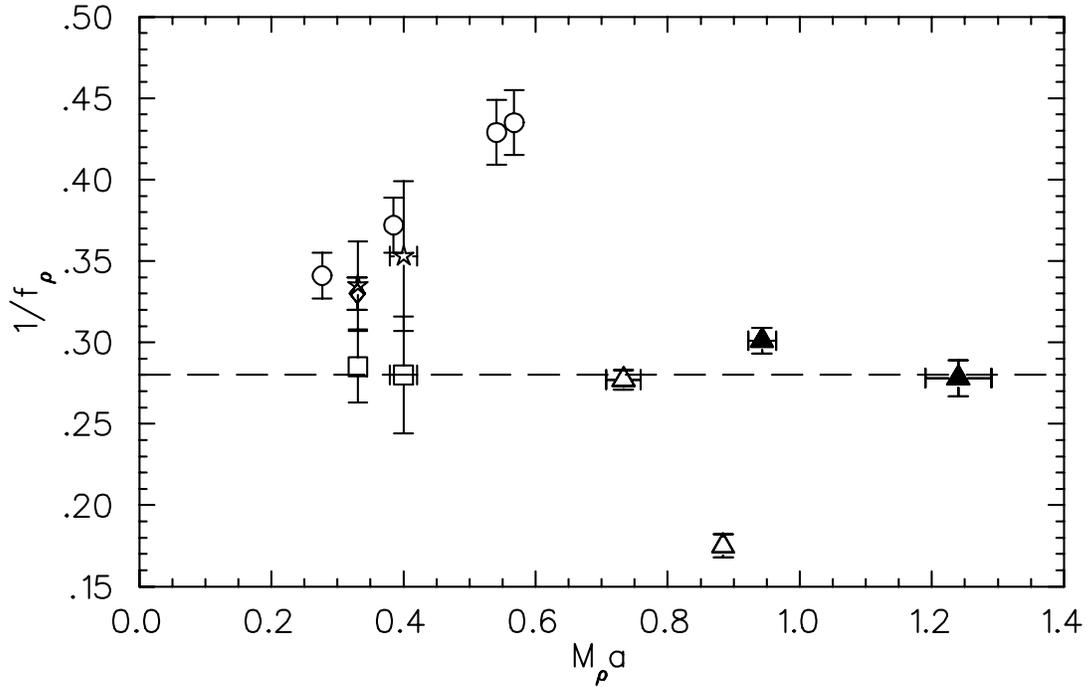}
\vspace{18pt}
\caption{The renormalized decay constant in the chiral limit.  All symbols
         are defined as in Fig.~\protect\ref{fig:ZfV}.  Squares and triangles
         use a nonperturbative determination of $Z_V$; all other data rely on
         a perturbative estimate.
         }\label{fig:fV}
\end{figure}
\begin{figure}[p]
\epsfxsize=380pt \epsfbox[30 419 498 732]{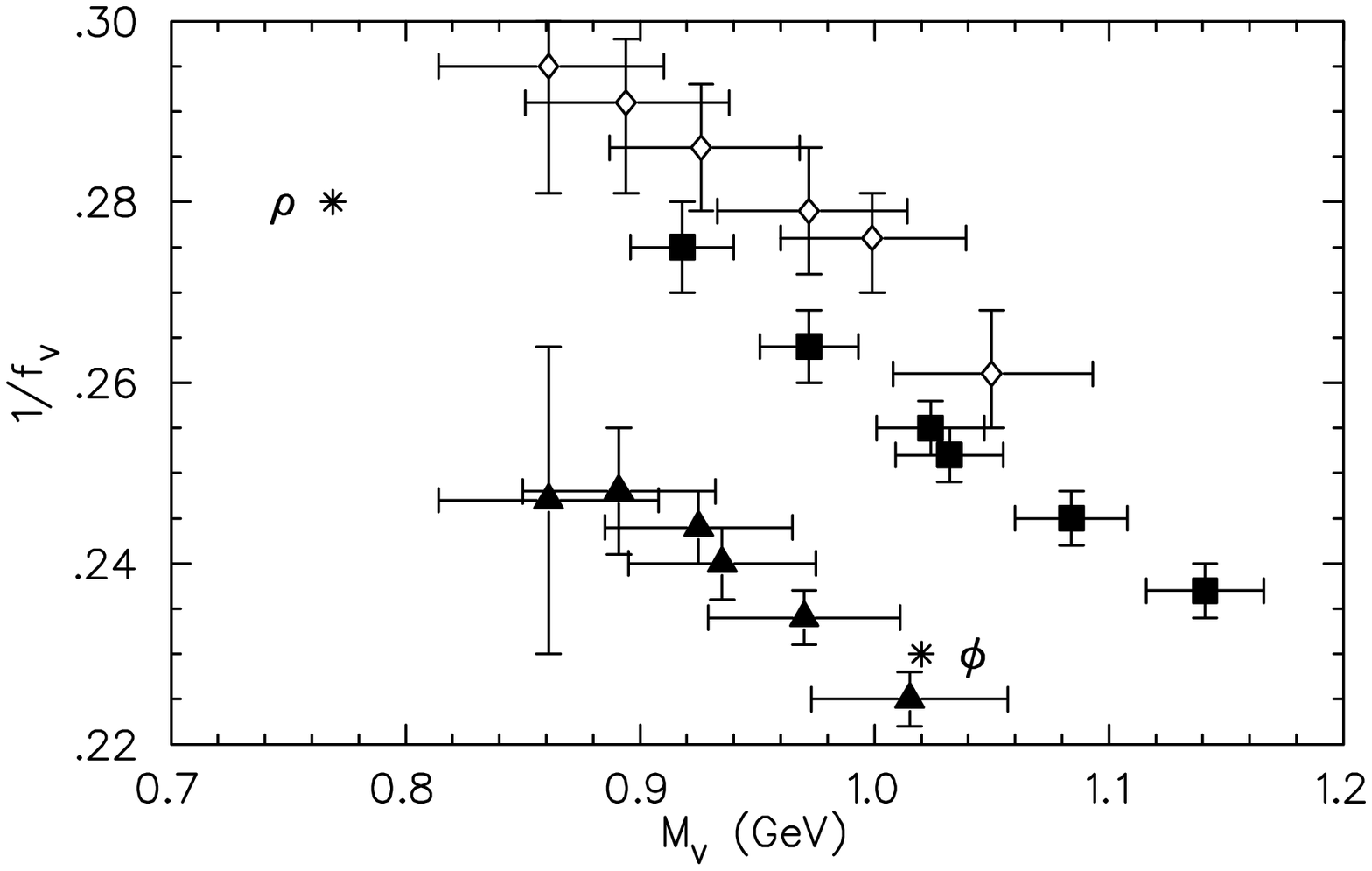}
\vspace{18pt}
\caption{Decay constant data before extrapolation/interpolation.  Solid 
         triangles(squares) are improved action data at $\beta=6.25$ (6.8),
         open diamonds are $\beta = 6.2$ clover data\protect\cite{UKQCD},
         and the asterisks represent experimental data.  The horizontal scale 
         is set by the physical $\rho$ meson mass.
         }\label{fig:fVraw}
\end{figure}
Our best determination of the $\rho$ meson
decay constant comes from the product of $Z_V$ (obtained from $R_D$)
and $1/(Z_Vf_\rho)$, which is shown in Fig.~\ref{fig:fV}.  The Wilson
calculations can be divided into two categories: those which compute $Z_V$
nonperturbatively as we have done in this work, and those which employ a
perturbative estimate of $Z_V$, denoted by $Z_{pert}$.  (The data in 
Fig.~\ref{fig:fV} with 
$Z_{pert}$ use quarks which are {\it not\/} normalized by the naive 
$\sqrt{2\kappa}$, see the original papers for details.)
The graph shows that both the improved and
Wilson actions give nonperturbative results which are quite independent of 
lattice spacing over a large range of $a$, and which are near the experimental 
value.
In contrast, the Wilson data using $Z_{pert}$ approaches a similar continuum
limit but with a noticeable dependence on $a$.

Having now discussed $1/f_V$ in the chiral limit, let's return to our
unextrapolated results, plotted in Fig.~\ref{fig:fVraw}.  Notice that the
slope of our data is consistent with the observed slope from the two
experimental points $1/f_\rho$ and $1/f_\phi$, but as $a \rightarrow 0$ 
the simulations appear to favour values larger than the experimental ones.
We find it interesting that the UKQCD data, also shown in Fig.~\ref{fig:fVraw},
displays the same tendency.  In their paper\cite{UKQCD}, the UKQCD
group concludes that this shift toward larger $1/f_V$ values is the opposite
of what would be expected from quenching,
and they suggest that finite volume effects or lattice spacing
effects may be responsible.  The testing of such possibilities within the
context of our ${\cal O}(a^2)$ classically-improved, tadpole-improved action
is much less computer-intensive than for the $\beta=6.2$ clover action, 
and might offer some valuable insight.

The hopping parameter corresponding to the strange quark as given in 
Table~\ref{tab:details}, was determined from the experimental relation
\begin{equation}
   \frac{m_{K^*}}{m_K} = 1.8
\end{equation}
by linear interpolation in $1/\kappa$ between the two neighbouring data points.
Using this, Table~\ref{tab:fV} gives
our results for the three light vector meson decay
constants and compares them to predictions from other groups.
\begin{table*}
\caption{The decay constants from various simulations.  Only the largest 
         $\beta$ results from each reference are shown here.
         All clover and Wilson data use the perturbative renormalization
         constant $Z_{pert}$.
         $a_\rho$ is the lattice spacing derived from the $\rho$ mass.
         }\label{tab:fV}
\begin{tabular}{lcccc}
\rule[-2mm]{0mm}{6mm} 
 Action & $a_\rho$[fm] & $1/f_\rho$ & $1/f_{K^*}$ & $1/f_\phi$ \\
\hline
 improved, $\beta=6.25$ & 0.32(1) & 0.278(11) & 0.255(6) & 0.236(2) \\
 improved, $\beta=6.8$ & 0.242(5) & 0.301(8) & 0.284(5) & 0.265(3) \\
 Wilson, $\beta=6.0$, Ref.~\protect\cite{BhaGup} & 0.0845(1) 
           & 0.33(1) & --- & --- \\
 Wilson, $\beta=6.17$, Ref.~\protect\cite{GF11} & 0.071(1) 
           & 0.34(1) & --- & 0.38(1) \\
 clover, $\beta=6.2$, Ref.~\protect\cite{UKQCD} & 0.073(3)
           & $0.316^{+7}_{-13}$ & $0.298^{+5}_{-9}$ & $0.280^{+3}_{-6}$ \\
 Wilson, $\beta=6.4$, Ref.~\protect\cite{AllGim} & ? 
           & 0.32(2) & 0.32(1) & --- \\
 clover, $\beta=6.4$, Ref.~\protect\cite{AllGim} & ? 
           & 0.25(2) & 0.24(1) & --- \\
 experiment & --- & 0.28 & --- & 0.23 \\
\end{tabular}
\end{table*}

\section{CONCLUSIONS}\label{Sec:conclude}

In this paper we have calculated vector current matrix elements and 
the renormalization
factor in the light quark sector for an improved action on lattices 
with spacings of about
0.3 to 0.4fm. This is the first step in going beyond the mass spectrum
with improved actions on coarse lattices.

The action used here is particularly simple. It has next-nearest-neighbour
couplings to remove ${\cal O}(a)$ and ${\cal O}(a^2)$ errors at the
classical level and also incorporates tadpole improvement.
No special tuning of parameters and no field transformations are required
with this action.

Vector meson decay constants and the vector current renormalization factor
calculated with the improved action were found to be comparable to those
with the Wilson action on lattices with about 1/3 the spacing. This is the
same as the improvement found previously in mass calculations and it
represents a very substantial saving in computer resources.

\acknowledgments

The authors are grateful to Peter Lepage for a helpful discussion.
This work was supported in part by the Natural Sciences and Engineering
Research Council of Canada.

\vfil\eject

\end{document}